\documentclass[preprintnumbers,nofootinbib,floatfix,twocolumn]{revtex4-2} 
\pdfoutput=1
\usepackage[english]{babel}
\usepackage{bm}
\usepackage{times}
\usepackage[normalem]{ulem}
\usepackage{hyperref}
\usepackage{xfrac}

\usepackage{graphicx,color}
\usepackage{colortbl, xcolor}
\hypersetup{
    linktocpage=true,
    setpagesize=true,
    colorlinks=true,
    urlcolor=blue,
    citecolor=blue,
    linkcolor=blue, 
    menucolor=blue,
    pdftitle={On the Role of LHC and HL-LHC at constraining Flavor Changing Neutral Currents},
    pdfsubject={Latex},
    pdfauthor={Kovalenko, Jesus, Zerwekh, Y Villamizar, Queiroz, Torres, Neto},
    pdfpagemode=FullScreen,
    pdfkeywords={BSM, FCNF, Meson Oscillations, Collider, LHC, FCC, HL, HE, 3-3-1 Models}
    }    
\usepackage{listings}
\usepackage{slashed}
\usepackage{appendix}
\usepackage{mathtools}
\usepackage{epsfig}
\usepackage{dcolumn}
\usepackage{textcomp}
\usepackage{appendix} 
\usepackage{multirow}
\usepackage{soul}
\usepackage{subfigure}
\usepackage[capitalize]{cleveref}
\usepackage{xspace}

\newcommand{\figs}[1]{\cref{#1}}
\newcommand{\eq}[1]{Eq.~(\ref{#1})}
\newcommand{\eqs}[1]{Eqs.~(\ref{#1})}
\newcommand{\tab}[1]{Table~\ref{#1}}
\newcommand{\zp}{Z^{\prime}}

\newcommand{\be}{\begin{equation}}
\newcommand{\ee}{\end{equation}}
\newcommand{\bi}{\begin{itemize}}
\newcommand{\ei}{\end{itemize}}

\begin{document}

\title{On the Role of LHC and HL-LHC in Constraining Flavor Changing Neutral Currents}

\author{A. S. de Jesus$^{4,5}$}
\email{alvarosndj@gmail.com}
\author{S. Kovalenko$^{2,3,8}$}
\email{sergey.kovalenko@unab.cl}
\author{T.~B.~de Melo$^{3,8}$}
\email{tessiomelo@institutosaphir.cl}
\author{J. P. Neto$^{4,5}$}
\email{jacintopaulosneto@gmail.com}
\author{Y.M. Oviedo-Torres}
\email{mauricio.nitti@gmail.com}
\author{F. S. Queiroz$^{3,4,5}$}
\email{farinaldo.queiroz@ufrn.br}
\author{Y. S. Villamizar$^{4,5,9}$}
\email{yoxarasv@gmail.com}
\author{A. R. Zerwekh $^{1,2,3}$}
\email{alfonso.zerwekh@usm.cl}

\affiliation{$^1$Departamento de F\'isica, Universidad T\'ecnica Federico Santa Mar\'ia, Casilla 110-V, Valparaiso, Chile.}
\affiliation{$^2$Centro Cient\'\i fico 	Tecnol\'ogico de Valpara\'\i so-CCTVal,	Universidad T\'ecnica Federico Santa Mar\'\i a, Casilla 110-V, Valpara\'\i so, Chile}
\affiliation{$^3$Millennium Institute for Subatomic Physics at the High-Energy Frontier (SAPHIR) of ANID, Fern\'andez Concha 700, Santiago, Chile}
\affiliation{$^4$International Institute of Physics, Universidade Federal do Rio Grande do Norte,
Campus Universitario, Lagoa Nova, Natal-RN 59078-970, Brazil}
\affiliation{$^{5}$Departamento de F\'isica, Universidade Federal do Rio Grande do Norte, 59078-970, Natal, RN, Brasil}
\affiliation{$^6$Departamento de Fisica, Universidade Federal da Paraiba, Caixa Postal 5008, 58051-970, Joao Pessoa, PB, Brazil}
\affiliation{$^7$Instituto de F\'isica de S\~ao Carlos, Universidade de S\~ao Paulo, Av. Trabalhador S\~ao-carlense 400, S\~ao Carlos-SP, 13566-590, Brasil.}
\affiliation{$^8$Center for Theoretical and Experimental Particle Physics - CTEPP,
Facultad de Ciencias Exactas, Universidad Andres Bello, Fernandez Concha 700, Santiago, Chile}

\affiliation{$^9$ Centro de Ci\^encias Naturais e Humanas, Universidade Federal do ABC, 09210-580, Santo Andr\'e, S\~ao Paulo, Brasil} 

\begin{abstract}
The Standard Model has no Flavor-Changing Neutral Current (FCNC) processes at the tree level. Therefore, processes featuring FCNC in new physics are tightly constrained by data. Typically, the lower bounds on the scale of new physics obtained from $K-\bar{K}$ or $B-\bar{B}$ mixing lie well above 10 TeV, surpassing the reach of current and future colliders. In this paper, we demonstrate, using a specific $\zp$ model that features flavor-changing interactions, that such limits can be severely weakened by specific choices of the quark mixing matrices with no prejudice while maintaining the CKM matrix in agreement with the data. We highlight the valuable role of the often-overlooked $D_0$ mixing in deriving robust FCNC limits and show that the LHC and HL-LHC are promising probes for flavor-changing interactions mediated by a $\zp$ boson.
\end{abstract}

\maketitle

\section{Introduction}

Flavor physics has been key in developing the Standard Model (SM). Indeed, most SM parameters are related to flavor physics, and thus can be determined using flavor-conserving and violating interactions. With the great experimental progress witnessed in the past decades, flavor physics became a powerful probe for new physics. It is often said that flavor physics can constrain very high-energy scales well beyond accelerators. Our work will revisit this narrative and show that, depending on the mixing used in the quark mixing matrices accelerators 
can be more constraining than flavor physics processes.

We do not mean to undervalue flavor physics. The underlying $SU(3)$ flavor symmetry of mesons and baryons motivated Gell-Mann to introduce the light quarks as the building blocks of hadronic matter \cite{Gell-Mann:1964ewy}. The existence of the charm quark was predicted to explain the smallness of the $K_L$ decay into muon pairs \cite{Glashow:1970gm}.  The need for a third generation and an estimation for the top quark mass were also related to flavor physics, particularly meson oscillations \cite{Buras:1993wr}. Consequently, flavor physics can never be overlooked.

We will be particularly interested in Flavor Changing Neutral Currents (FCNCs) in our work. FCNC is a process in which the initial and final states have the same electric charge but a different flavor configuration, for instance, a process in which the initial state has a d quark, but the final state has a $s$ quark, instead. Hence, for a flavor transition to occur in the SM, the exchange of at least a W boson is required. Such processes have a reduced rate relative to flavor-conserving interactions because of the Glashow-
Iliopoulos-Maiani (GIM) mechanism and because FCNCs appear only at the loop level. 
For this same reason, new physics models can be probed up to very high-energy scales, often well beyond the reach of past and current high-energy colliders \cite{Buras:2013ooa}, but this declaration is not always valid, as we will show.

\begin{figure}
    \centering
    \includegraphics[width=0.9\columnwidth]{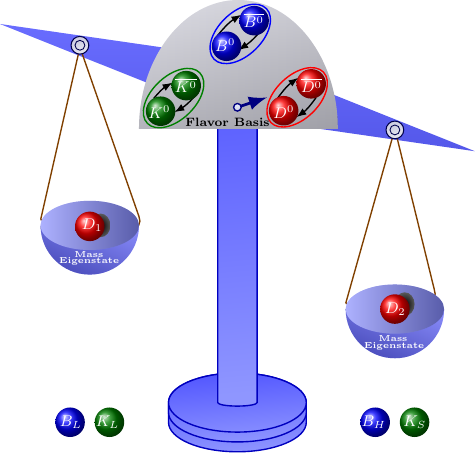}%
    \caption{How flavor eigenstates of mesons, lead to mass eigenstates of mesons with different masses.}
    \label{fig1}
\end{figure}

Our reasoning will not be general, though. It will target FCNCs that contribute to meson mixings arising from a $\zp$ boson. Mesons are made up of one quark and one antiquark. For example, the meson $D_0$ has a charm quark and an up antiquark. Given the small mass difference between $D_0$ and $\bar{D_0}$, these particles oscillate into one another. In other words, flavor-changing interactions are present at the one-loop level. A similar pattern appears in other meson systems, namely $K^{0}-\bar{K}^{0}$, $B^0_d-\bar{B^0}_d$ and $B^0_s-\bar{B^0}_s$. In Figure \ref{fig1}, a diagram illustrates these systems. The mass differences between these mesons have been measured with great accuracy. Therefore, any new physics effect that could contribute to them is tightly constrained by data. This conclusion has been extensively explored in the literature. In extra gauge symmetries where the $\zp$ field interacts differently between fermion generations, meson mixings depend on the quark mixing matrices \cite{CarcamoHernandez:2022fvl}. The product of the quark mixing matrices results in the measured CKM matrix. One cannot arbitrarily change the quark mixing matrices because the CKM matrix must be reproduced. However, FCNC processes mediated by a $\zp$ field depend on the product of individual entries of the quark mixing matrices, allowing one to modify their values and suppress the $\zp$ contribution to FCNCs while keeping the CKM matrix in agreement with the data.

We will address this issue in a particular extension of the SM based on a $SU(3)_c \otimes SU(3)_L \otimes U(1)_N$ gauge group, typically referred to in the literature as 3-3-1 model \cite{Pisano:1991ee,Foot:1992rh,Montero:1992jk,Dias:2009au}. FCNC studies have been done for different particle contents having the underling $SU(3)_c \otimes SU(3)_L \otimes U(1)_N$ gauge group \cite{GomezDumm:1994tz,Long:1999ij,Rodriguez:2004mw,CarcamoHernandez:2005ka,Palcu:2007um,Promberger:2007py, Dong:2008sw, Dong:2008sw,Cabarcas:2009vb,Benavides:2009cn,Dong:2010zu, Dong:2011vb, Cogollo:2012ek,Cogollo:2013mga,Giang:2012vs, Hue:2013uw,Machado:2013jca, Hue:2013uw, CarcamoHernandez:2013krw,RebelloTeles:2013dkw,Vien:2014ica, Vien:2014pta, Hue:2015fbb, Hue:2015fbb,CarcamoHernandez:2015rmj,Vien:2015wca,Queiroz:2016gif,Buras:2016dxz,Machado:2016jzb,Fonseca:2016tbn,Singh:2018zmx,Singh:2019hvj,Huong:2019vej, Hong:2020qxc,Huong:2020csh,NguyenTuan:2020xls,Oliveira:2022vjo,Oliveira:2022dav,CarcamoHernandez:2022fvl,Buras:2023fhi,Buras:2023ldz}. It can be concluded that the bounds on the mass of the $\zp$ derived from those studies lie above the 10 TeV scale, making those gauge bosons inaccessible at the LHC and High-Luminosity-LHC accelerators. The goal of our work differs from previous studies because we aim to demonstrate that the LHC and High-Luminosity LHC are promising colliders capable of discovering $\zp$ bosons that mediate FCNC in agreement to FCNC bounds.

In particular, we will conclude that one can choose the quark mixing matrices in such a way that FCNC bounds on the $\zp$ mass range between 0.1-3TeV, rendering LHC and HL-LHC much more promising probes for the $\zp$ bosons from 3-3-1 symmetry. Our work is structured as follows: in section \ref{RFCNC}, we review FCNC in the SM; in section \ref{EF}, we discuss a simplified model approach to FCNC; in section \ref{IV}, we address FCNC in a 3-3-1 model; in section \ref{V}, we discuss our findings and put them in perspective with collider physics; in section \ref{VI}, we draw our conclusions.

\begin{figure*}
    \centering
    \subfigure{\includegraphics[scale=0.16]{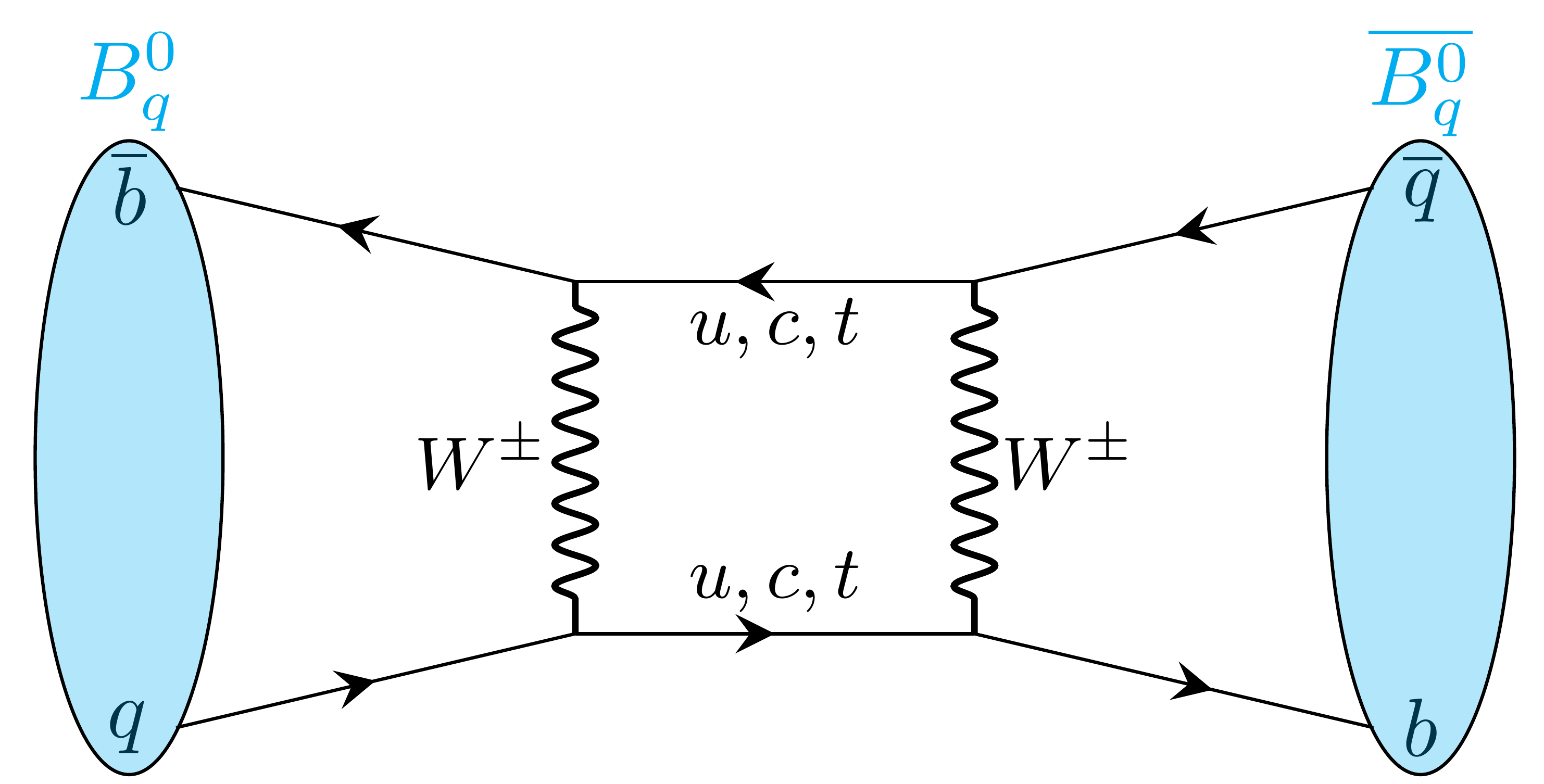}}
    \subfigure{\includegraphics[scale=0.16]{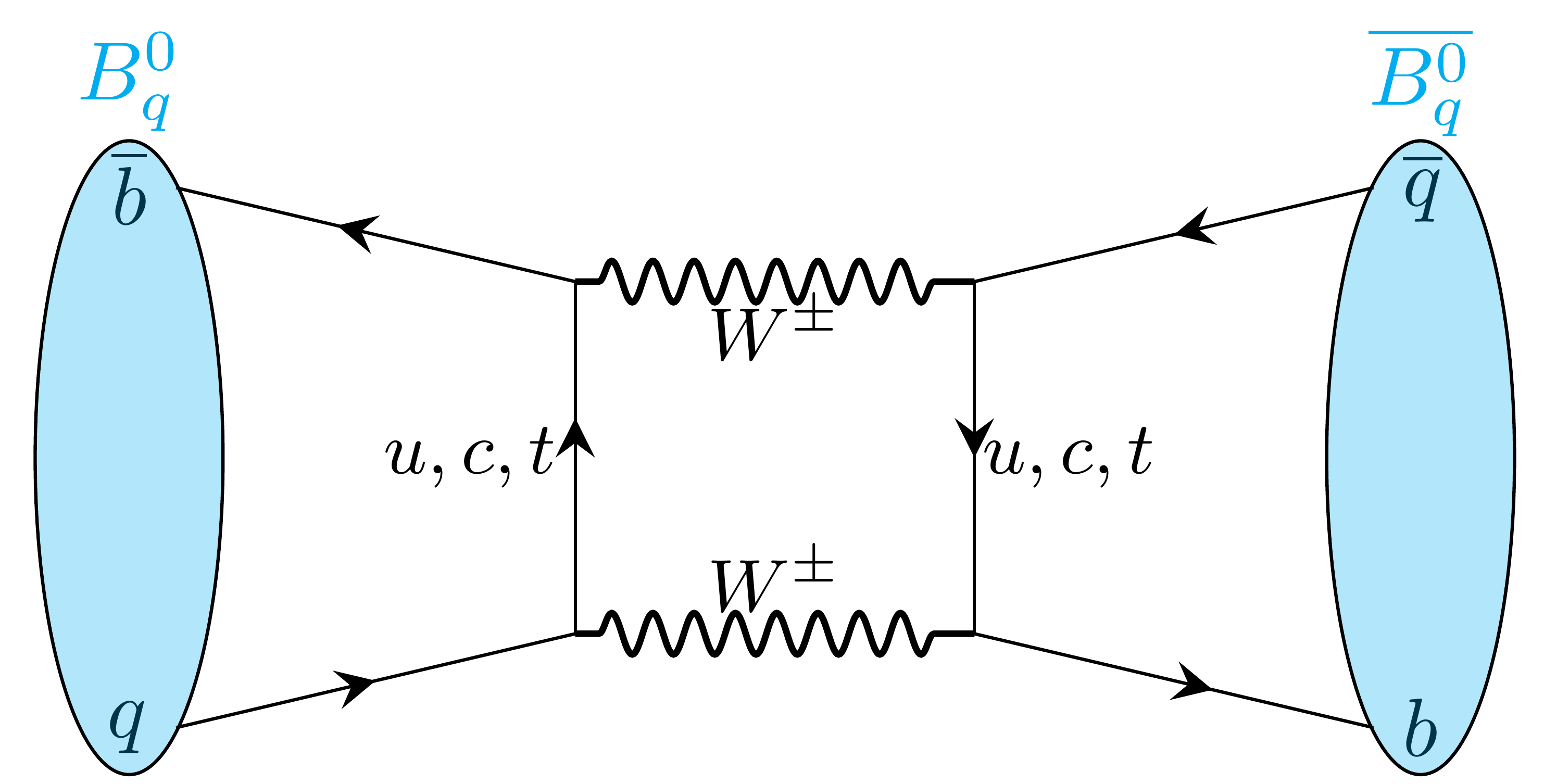}}
    \subfigure{\includegraphics[scale=0.16]{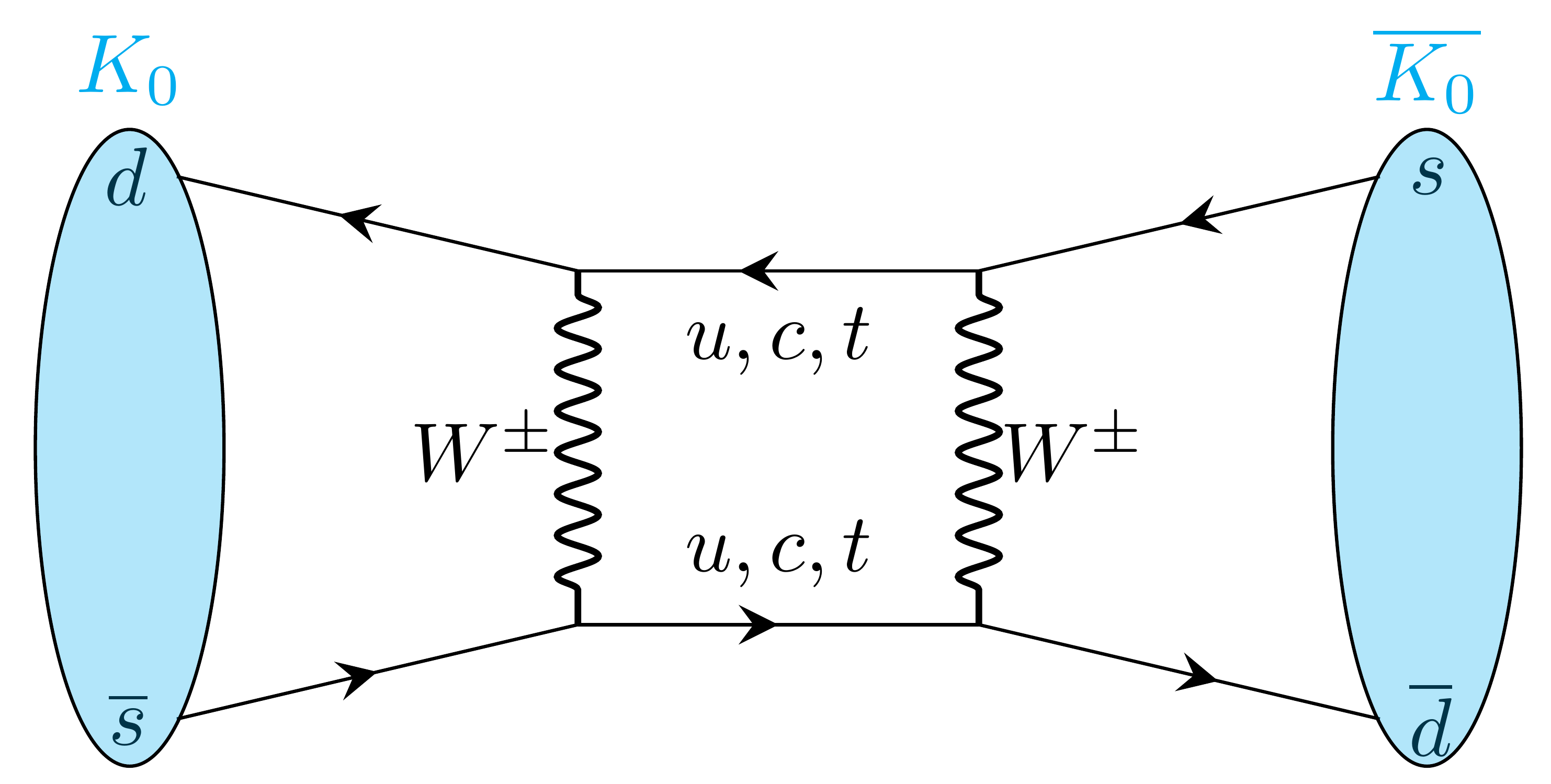}}
    \subfigure{\includegraphics[scale=0.16]{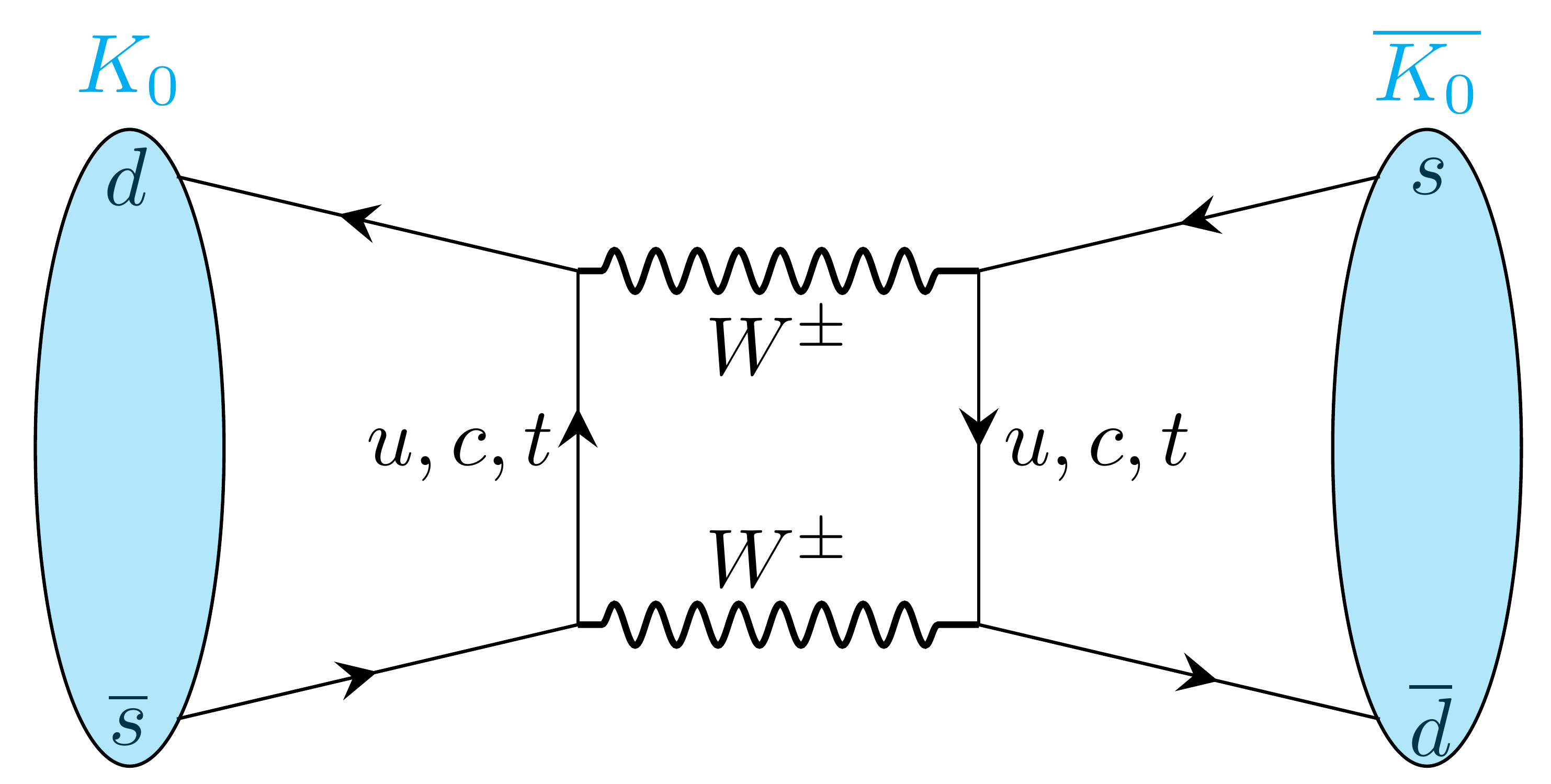}}
    \subfigure{\includegraphics[scale=0.16]{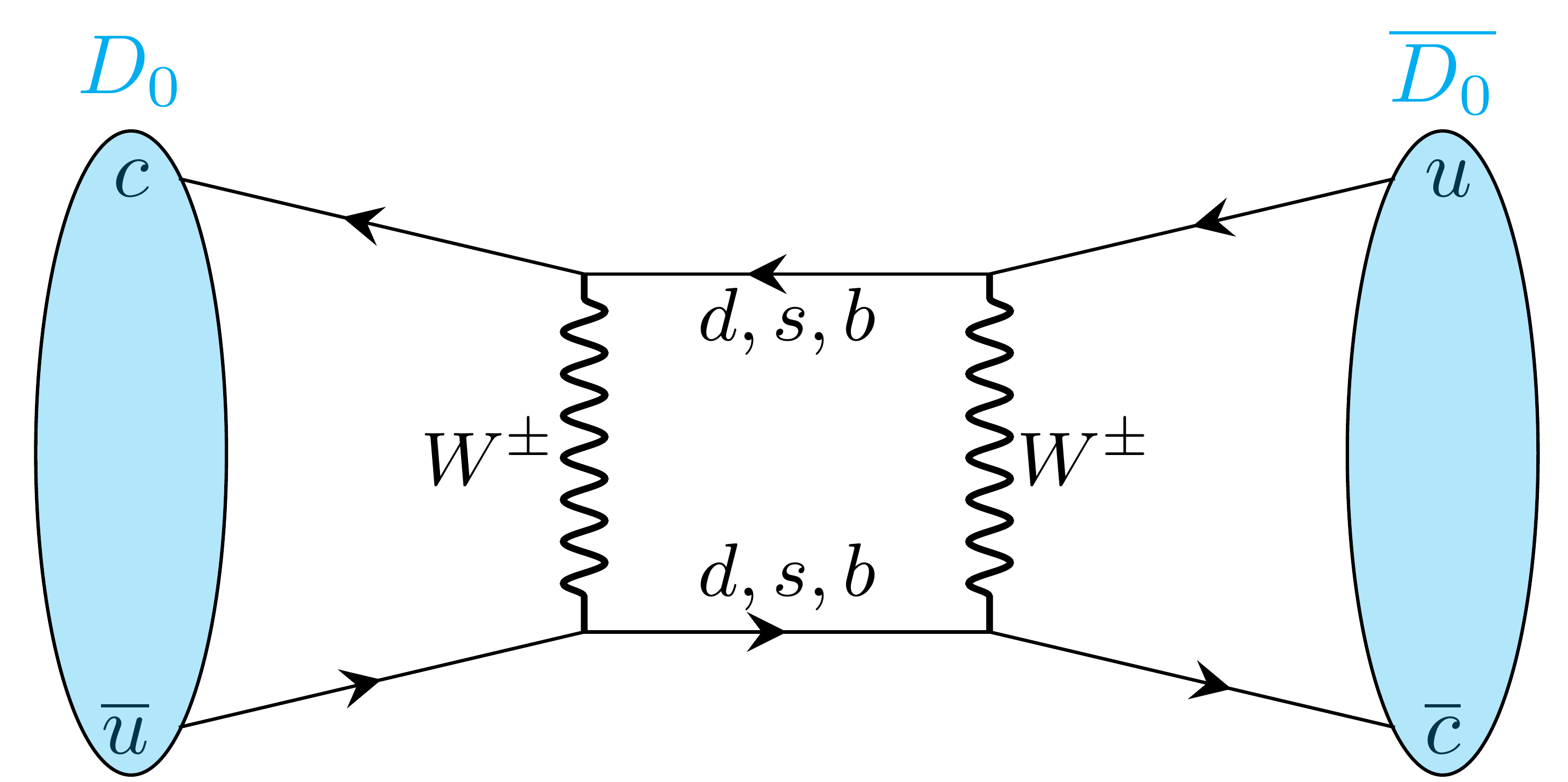}}
    \subfigure{\includegraphics[scale=0.16]{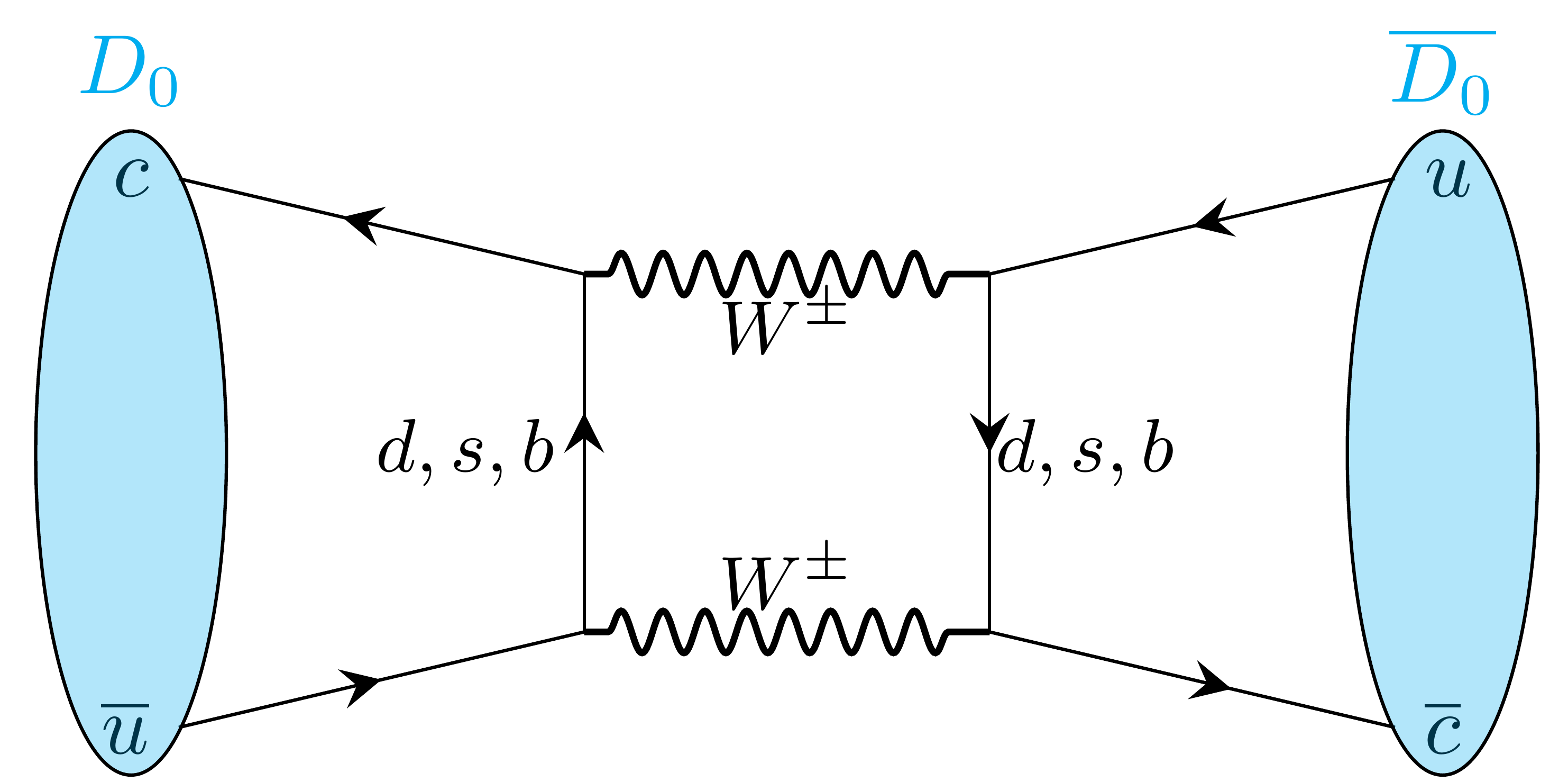}}
    \caption{Dominant box Feynman diagrams that contribute to flavor-changing processes of the four meson systems under investigation: (a) and (b) $B_q^0 - \overline{B_q^0}$ mixing (with $q=d$ or $s$); (c) and (d)  $K_0 - \overline{K}_0$ mixing; (e) and (f) $D_0 - \overline{D}_0$ mixing.}
    \label{fig2}
\end{figure*}

\section{\label{RFCNC} Review of Flavor Changing Neutral Current}

Before discussing how FCNC emerges in a $\zp$ model, for pedagogical reasons, we will review the key ingredients of FCNC in the SM. Later, we will address why there is no FCNC at the tree level.

\subsection{How FCNCs Emerge in the SM}

We will introduce the SM and show why it arises at the loop level. We remind the reader that the left-handed quarks are arranged as $SU(2)_L$ doublets,
\be\label{eq:quarks}
Q_j =
\begin{pmatrix}
u_L \\ d_L
\end{pmatrix},
\begin{pmatrix}
c_L \\ s_L
\end{pmatrix},
\begin{pmatrix}
t_L \\ b_L
\end{pmatrix}\,,
\ee
while the right-handed quarks are $SU(2)_L$ singlets,
\be
U_j=u_R,c_R,t_R \qquad D_j=d_R,s_R,b_R \,.
\ee

The relevant terms for our discussion are the charged and neutral currents and the Yukawa Lagrangian. The charged and neutral currents arise from the kinetic term,

\be\label{eq:Lfermion}
\mathcal{L}_\text{fermion} = \sum_{j=1}^3
\bar Q_j i \slashed D_Q Q_j + \bar U_j i \slashed D_U U_j + \bar D_J i \slashed D_D D_j \vspace{-3mm}
\,,
\ee
with the index $j$ running over the three generations of quarks. The covariant derivatives are defined as,
\begin{eqnarray}
D_{Q,\mu} &=& \partial_\mu + i g_s T^a G^a_\mu + i g \tau^a W^a_\mu + i g' Q^Y_Q B_\mu\,,
\nonumber\\
D_{U,\mu} &=& \partial_\mu + i g_s T^a G^a_\mu \hspace{1.8cm} +  i g' Q^Y_U B_\mu\,,
\nonumber\\
D_{D,\mu} &=& \partial_\mu + i g_s T^a G^a_\mu \hspace{1.8cm}  +  i g' Q^Y_D B_\mu\,,\label{eq:DDmu}
\end{eqnarray}
with the hypercharges $Q^Y_Q = 1/6$, $Q^Y_U = 2/3$, $Q^Y_D = -1/3$, and $\tau^a (a=1,2,3)$, $T^a (a=1,\dots,8)$ being the generators of the $SU(2)_L$ and $SU(3)_c$ groups, respectively. Notice that the same gauge coupling appears in front of the $G_\mu^a$, $W_\mu^a$, and $B_\mu$ fields, rendering their respective interactions flavor universal across all three generations. However, flavor non-universality is introduced in the Yukawa Lagrangian, 

\be\label{eq:Yuk}
\mathcal{L}_\text{Yuk} = \sum_{i,j=1}^3 (
- Y_{U,ij} \bar Q_{Li} \tilde{H} U_{Rj} - Y_{D,ij} \bar Q_{Li} H D_{Rj}  + h.c.)\,,
\ee
where $H$ is the Higgs doublet, and $h.c.$ abbreviates the hermitian conjugate term. Note that $i,j$ are generation indices and $\tilde H = \epsilon H^* = (H^{0*},-H^-)^T$. After spontaneous symmetry breaking,  in the unitary gauge, we take $\langle H \rangle = (0,v)^T$, and obtain the quark masses,
\be
\sum_{i,j=1}^3 (
- m_{U,ij} \bar u_{Li} u_{Rj} - m_{D,ij} \bar d_{Li}  d_{Rj} + h.c.)\,.
\ee

As $i,j$ are generation indices, $m_{U,D}$ can be seen as $3 \times 3$ matrices in flavor space with $m_{U,D} = v Y_{U,D}$. We need to diagonalize them to derive the quark masses and the mass eigenstates ($u^\prime,d^\prime$), using a bi-unitary transformation as follows,

\begin{eqnarray}
u_L &=&  V_L^U u_L^\prime\,,\qquad u_R =  V_R^U u_R^\prime \,, \nonumber\\
\qquad &d_L =&  V_L^D d_L^\prime \,, \qquad d_R =  V_R^D d_R^\prime \,.
\label{eqtransformations}
\end{eqnarray}
where $V^{U}_L$,$V^{U}_R$, $V^{D}_L$,$V^{D}_R$ are the rotation matrices that relate the flavor fields with the mass eigenstates. The charged current resulted from Eq.\ref{eq:DDmu} is,

\begin{equation}
\mathcal{L} = -\frac{g}{\sqrt{2}}  \bar u_{Li} \gamma_\mu d_{Li} W^{\mu+} .
\label{eqchargedcurrent}
\end{equation} 

Applying Eq. \ref{eqtransformations} in Eq.\ref{eqchargedcurrent} we get, 

\be
\mathcal{L} =-\frac{g}{\sqrt{2}}  \bar u_{Li}^\prime  \left(  V_{L,ij}^{U\dagger}  V_{L,jk}^D \right) \gamma_\mu d_{Lk}^\prime\,  W^{\mu+} ,
\ee which gives rise to the CKM matrix \cite{Cabibbo:1963yz,Kobayashi:1973fv},

\be
 V_\text{CKM} =  V_L^{U\dagger}  V_L^{D}. \label{VCKM}
\ee

Hence, in the SM, we do have flavor-changing interactions at tree level, which are mediated by the W boson. However, they do not represent FCNC. It is important to note that the Cabibbo-Kobayashi-Maskawa (CKM) matrix plays a key role in FCNC, and it governs the $B_d$, $B_s$, $K_0$, and $D_0$ mixings shown in Fig.\ref{fig2}. The CKM matrix is often parameterized as follows,

\be
 V_\text{CKM} =
\begin{pmatrix}V_{ud}&V_{us}&V_{ub}\\
V_{cd}&V_{cs}&V_{cb}\\V_{td}&V_{ts}&V_{tb}\end{pmatrix}\,.
\label{eqCKM}
\ee and its entries are measured to be \cite{Workman:2022ynf},

\begin{align}
    \label{eqCKM2}
&\left|V_{\mathrm{CKM}}\right|= \\
&\left(
\begin{array}{ccc}
0.97435  \scriptstyle{\pm 0.00016} & 0.22500  \scriptstyle{\pm 0.00067} & 0.00369  \scriptstyle{\pm 0.00011} \\
0.22486  \scriptstyle{\pm 0.00067} & 0.97349  \scriptstyle{\pm 0.00016} & 0.04182_{-0.000074}^{+0.00085} \\
0.00857_{-0.00018}^{+0.00020} & 0.04110_{-0.00072}^{+0.00083} & 0.999118_{-0.000036}^{+0.000031}
 \end{array}\right). \nonumber
\end{align} 

Moreover, the CKM matrix is unitary. Therefore, the unitarity of the CKM matrix implies that the sum of the products of the elements in each row and column should be zero. Thus, the expression
\begin{equation}
\sum_{i=u,c,t} V^\ast_{is}V_{ib} = 0, 
\label{equnitarity}
\end{equation}
represents a sum over the up-type quarks ($u, c$, and $t$) in the CKM matrix. It is part of the unitarity relation of the CKM matrix, which ensures the conservation of total probability for transitions between different quark flavors in weak interactions.

Any FCNC study should make sure that \cref{eqCKM,eqCKM2,equnitarity} are respected. We have noticed that flavor-changing interactions arise in processes mediated by the W boson. Now it is time to explain why FCNC is absent in the SM at the tree level.

\subsection{Absence of FCNCs in the SM}\label{sec:FCNC}

 To see how FCNCs are absent at the tree level, we need to revisit the Z boson interaction with left-handed quarks, which is found to be,

\begin{equation}
\mathcal{L}_Z = \frac{g}{\cos\theta_W}\left(-\frac{1}{2}+\frac{1}{3}\sin^2\theta_W\right) \left( \bar d_{Lj}  \gamma_\mu  d_{Lj} \right)  Z^{\mu}.
\end{equation}

Similarly to what we did in the charged current, we need to apply the transformation defined in Eq.\ref{eqtransformations} to obtain the Z interaction with mass eigenstate quarks,

\begin{eqnarray}
\mathcal{L}_Z = \frac{g}{\cos\theta_W}\left(-\frac{1}{2}+\frac{1}{3}\sin^2\theta_W\right) \times \nonumber\\
\left[ \bar d_{Li}^\prime (V^{D\dagger}_L)_{ij}(V^D_L)_{jk}  \gamma_\mu  d_{Lk}^\prime \right] Z^{\mu}
\end{eqnarray}which simplifies to,

\begin{equation}
\mathcal{L}_Z= \frac{g}{\cos\theta_W}\left(-\frac{1}{2}+\frac{1}{3}\sin^2\theta_W\right) \bar d_{Li}^\prime \delta_{ik} \gamma_\mu   d_{Lk}^\prime Z^{\mu},
\end{equation}

Therefore, the unitarity of the flavor rotation matrix, $V^D_L$, which implies in $(V^{D\dagger}_L)_{ij}(V^D_L)_{jk}=\delta_{ik}$, makes the neutral current flavor diagonal. This conclusion holds for all fermions. Thus, FCNC is absent in the SM at the tree level, but we have noticed that it surfaces at one loop level through the diagrams displayed in Fig.\ref{fig2}. Besides, in the SM, right-handed quarks do not participate in FCNCs due to the suppression of their interaction with the charged $W$ bosons. This is because of the violation of parity, which makes the weak interactions different for right-handed and left-handed quarks. Instead, right-handed quarks primarily interact through the neutral $Z$ boson, which does not allow flavor changes. After this short review of FCNC, we will address FCNC using an effective field theory approach.

\begin{table}
{\color{blue} Input parameters}\\
\begin{tabular}{|c|}
\hline
$\left(\Delta m_K\right)_{exp}=\left(3.484\pm 0.006\right)\times 10^{-12}$ MeV \\
$\left(\Delta m_K\right)_{SM}=3.483\times 10^{-12}$ MeV \\
 $m_K=\left(497.611\pm 0.013\right)$ MeV  \\
 $\sqrt{B_K}f_K=131$ MeV \\
 $\eta_K=0.57$ \\
\hline

\hline
$\left(\Delta m_D\right)_{exp}=\left(6.25316_{-2.8962}^{+2.69873}\right)\times 10^{-12}$~MeV  \\
$\left(\Delta m_D\right)_{SM}=10^{-14}$~MeV  \\
$m_D=\left(1865\pm 0.005\right)$ MeV  \\
$\sqrt{B_D}f_D=187$ MeV  \\
$\eta_D=0.57$  \\
\hline

\hline
$\left(\Delta m_{B_d}\right)_{exp} =\left(3.334\pm 0.013\right) \times 10^{-10}$ MeV \\ 
$\left(\Delta m_{B_d}\right)_{SM} =\left(3.653\pm 0.037\pm 0.019\right) \times 10^{-10}$ MeV \\
 $m_{B_d}=\left(5279.65\pm 0.12\right)$ MeV \\
$\sqrt{B_{Bd}}f_{Bd}=210.6$ MeV \\
$\eta_{B_d}=0.55$ \\
\hline

\hline
$\left(\Delta m_{B_s}\right)_{exp} =\left(1.1683\pm 0.0013\right) \times 10^{-8}$ MeV \\ 
$\left(\Delta m_{B_s}\right)_{SM} =\left(1.1577\pm 0.022\pm 0.051\right) \times 10^{-8}$ MeV \\
 $m_{B_s}=\left(5366.9\pm 0.12\right)$ MeV \\
$\sqrt{B_{B_s}}f_{B_s}=256.1$ MeV  \\
$\eta_{B_s}=0.55$ \\
\hline
\end{tabular}

\caption{Meson masses \cite{CPLEAR:1998zfe,Artuso:2015swg,Jubb:2016mvq,Wang:2018csg,Lenz:2019lvd,HFLAV:2019otj,Aoki:2021kgd} and the values of the bag parameters \cite{Workman:2022ynf,Aoki:2021kgd}.}
\label{tableI}
\end{table}

\section{\label{EF}FCNC and Simplified Model}

In this section, we intend not to provide a comprehensive review of FCNC using an effective field theory. For readers interested in such a review, we suggest referring to the works of \cite{Buras:1998raa,Buras:2005xt,Grossman:2010gw,Isidori:2013ez}. For correlations with recent flavor anomalies, see \cite{Ray:2022bxg}. Our FCNC approach will be limited to
meson oscillations. Interestingly, the neutral mesons $K$, $D$, $B_d$, and $B_s$ are the only hadrons that mix with their antiparticles. These meson states $K (\bar{s}d)$, $D(c\bar{u})$, $B_d (\bar{b} d)$, $B_s(\bar{b} s)$ and their antiparticle oscillate into each other, and the mass difference between them has been measured. Many $B_d-\bar{B}_d$ oscillations have been measured, which led to a precise measurement of the mass difference \cite{ALEPH:1996ofb,DELPHI:1997pdr,LHCb:2013fep,LHCb:2016gsk}. As for $B_s -\bar{B}_s$ mixing, it was firstly observed at LEP \cite{CDF:2006imy}, and later at LHCb detector \cite{LHCb:2011vae,LHCb:2013lrq}. 
The first evidence for $D_0-\bar{D}_0$ mixing was seen in 2007 by Belle \cite{BELLE:2007dgh} and BaBar \cite{BaBar:2007kib}. These findings were ratified by CDF collaboration \cite{CDF:2007bdz} and LHCb \cite{LHCb:2012zll}. With the numerous measurements of the meson mixings the precision has greatly improved, allowing one to constrain new physics effects \cite{ParticleDataGroup:1986kuw}.

\begin{figure}
    \centering
    \subfigure{\includegraphics[scale=0.14]{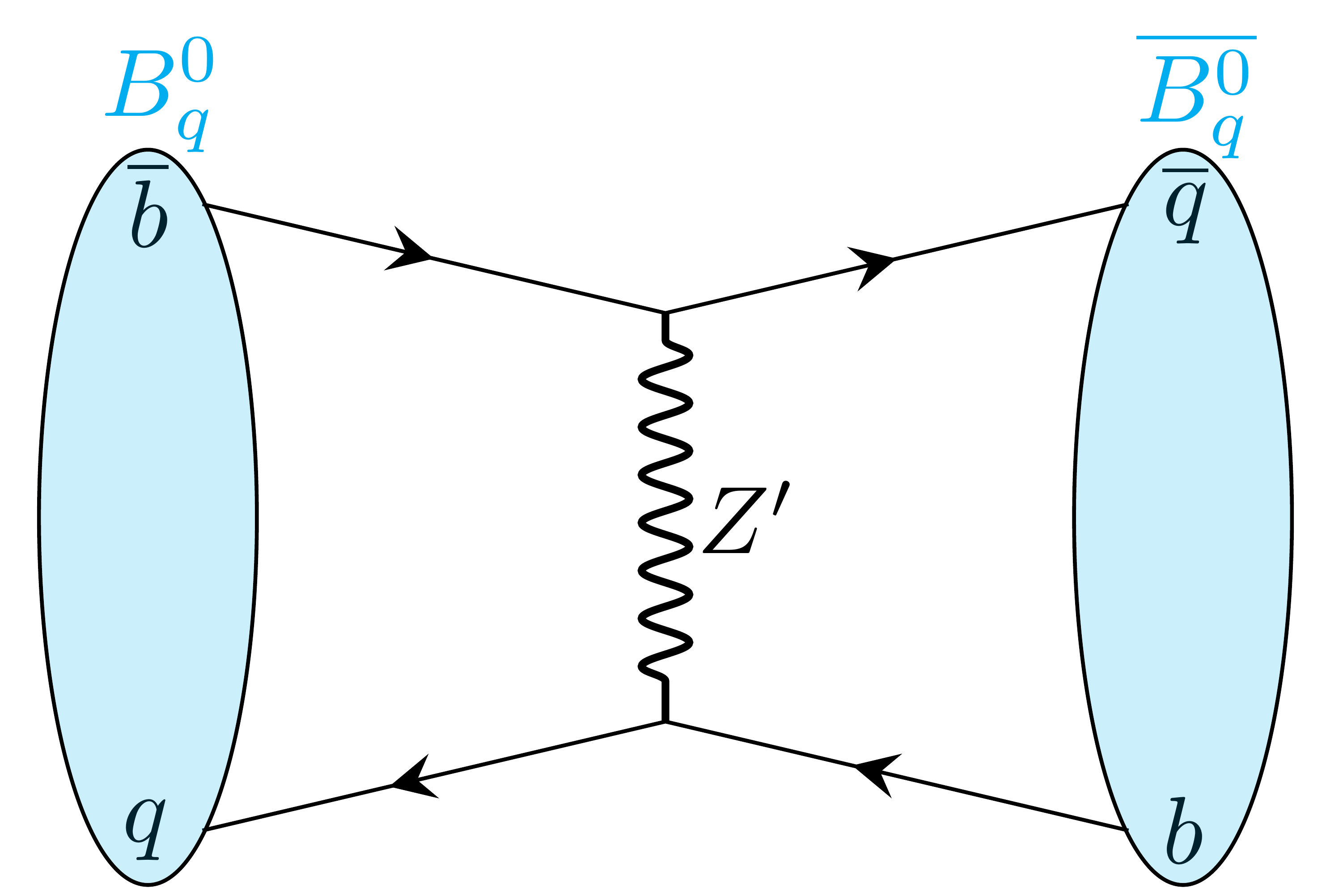}}
    \subfigure{\includegraphics[scale=0.14]{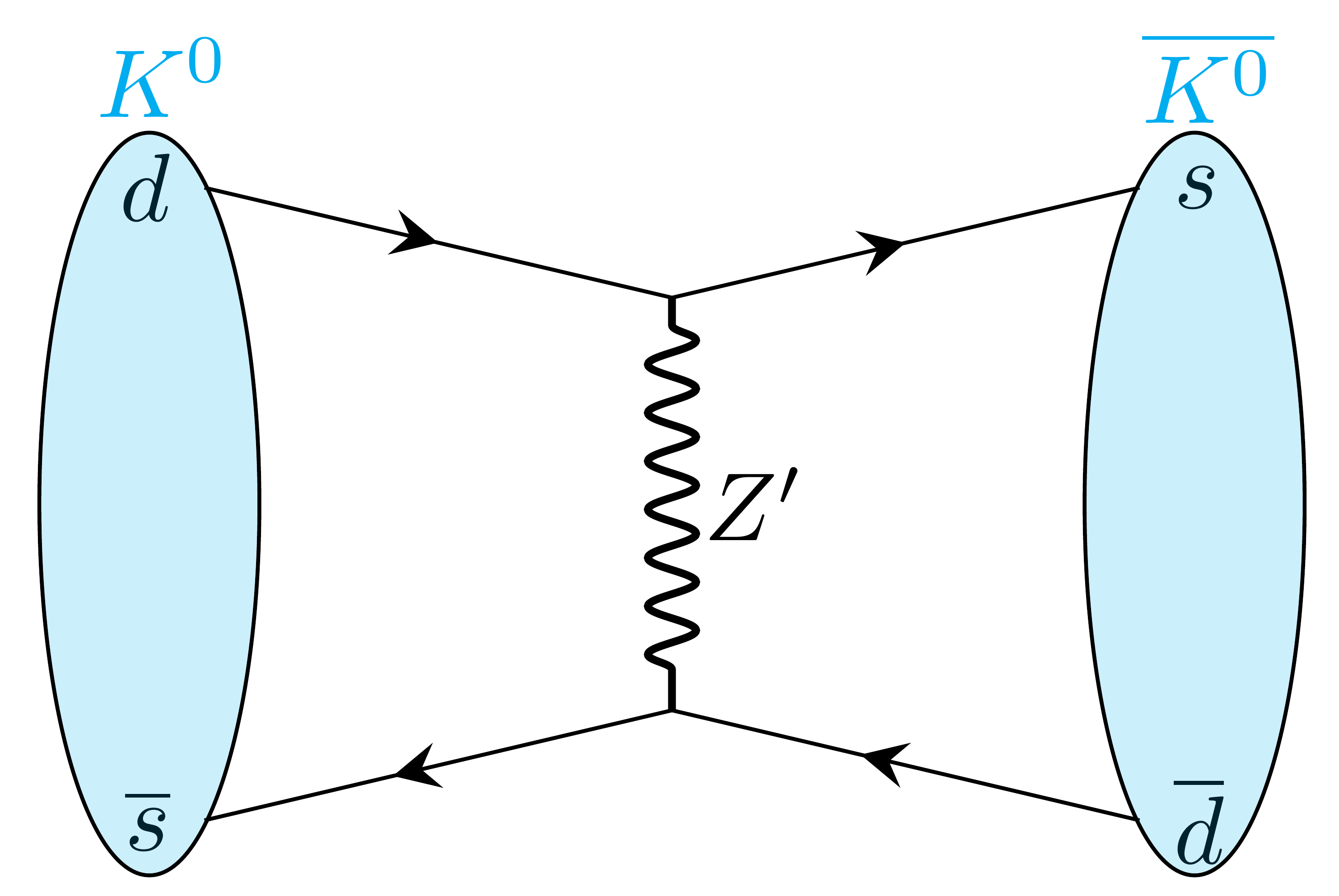}}
    \subfigure{\includegraphics[scale=0.14]{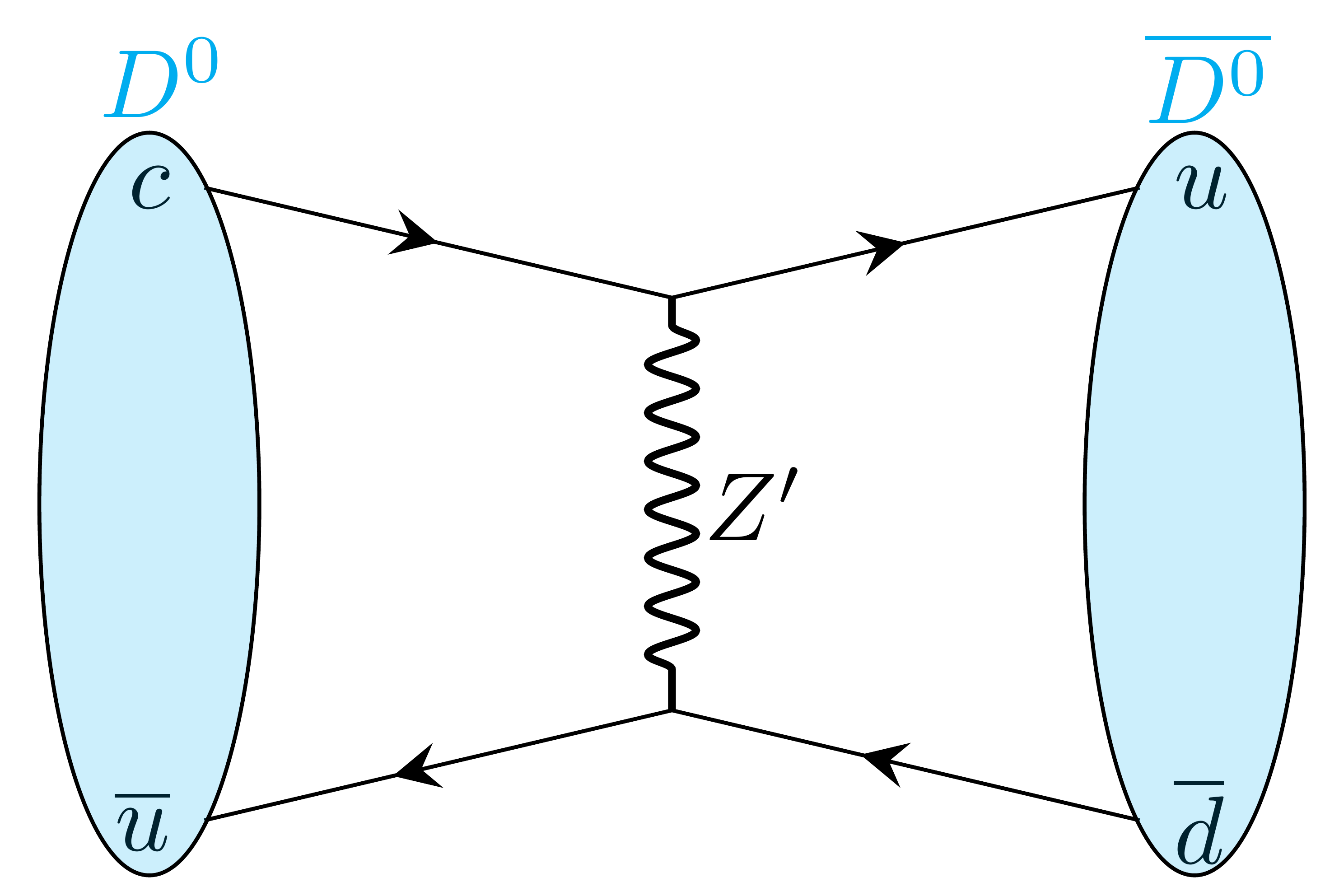}}
    \caption{Feynman diagrams at tree level illustrating the contribution of the $\zp$ gauge boson in flavor-changing processes of the four meson systems under investigation: (a) $B_q^0 \rightarrow{\overline{B_q^0}}$ transitions ($q=d$ or $s$); (b)  $K_0 \rightarrow{\overline{K}}_0$ transitions; (c) $D_0 \rightarrow{\overline{D}}_0$ transitions.}
    \label{fig3}
\end{figure}

That said, our effective field theory approach focuses on interactions mediated by a $\zp$, such as those present in 3-3-1 models. Specifically, we consider a $\zp$ that distinguishes one fermion generation from the others. The mass difference of meson systems can be impacted by sizeable flavor transitions caused by heavy gauge and $\zp$ boson see Fig.~\ref{fig3}. With this in mind, we can write down an effective Lagrangian that assumes the new physics interacts equally with the first two generations but differently with the third generation, as follows:

\begin{eqnarray}
\ensuremath{\mathcal{L}}^{\zp}_{q}  & \supset & g^{\prime} \bar{q}_{k L} \gamma_\mu q_{k L}+g^{\prime \prime} \bar{q}_{3 L} \gamma_\mu q_{3 L},
\label{eq:FCNC_new} 
\end{eqnarray}

with $g^{\prime}$  and $g^{\prime \prime}$ 
 represent the coupling constants between the $\zp$ boson and $u$-type and $d$-type quarks (being $k=1,\, 2,$, running through the two fermion generations) and the third fermion generation, respectively. Note that 
 Eq. \eqref{eq:FCNC_new} is in the flavor state basis, and once we rotate to the mass eigenstate basis, FCNC arises. The mass eigenstate and flavor bases are connected through Eq.\ref{eqtransformations}. After rotation, we find the effective Lagrangians relevant for the meson mixing,
\begin{eqnarray}
 \mathcal{L}^{K_0-\bar{K}_0}_{\zp \, eff} &=& \frac{1}{\Gamma^{ 2}}|(V_{L}^{D})_{31}^*(V_{L}^{D})_{32}|^2|\bar{d}^{\prime}_{1L}\gamma_\mu d^{\prime}_{2L}|^2, \nonumber\\
 \mathcal{L}^{D_0-\bar{D}_0}_{\zp \, eff} &=& \frac{1}{\Gamma^{2}}|(V_{L}^{U})_{31}^*(V_{L}^{U})_{32}|^2|\bar{u}^{\prime}_{1L}\gamma_\mu u^{\prime}_{2L}|^2, \nonumber\\
 \mathcal{L}^{B^0_d-\bar{B}^0_d}_{\zp \, eff} &=& \frac{1}{\Gamma^{2}}|(V_{L}^{D})_{31}^*(V_{L}^{D})_{33}|^2|\bar{d}^{\prime}_{1L}\gamma_\mu d^{\prime}_{3L}|^2,\nonumber\\
 \mathcal{L}^{B^0_s-\bar{B}^0_s}_{\zp \, eff} &=& \frac{1}{\Gamma^{2}}|(V_{L}^{D})_{32}^*(V_{L}^{D})_{33}|^2 |\bar{d}^{\prime}_{2L}\gamma_\mu d^{\prime}_{3L}|^2,\nonumber
 \label{eq:FCNC8}
\end{eqnarray}and consequently,
\begin{eqnarray}
 (\Delta m_K)_\Gamma &=& \frac{1}{\Gamma^2} |(V_{L}^{D})_{31}^*(V_{L}^{D})_{32}|^2 f_K^2 B_K \eta_K m_K, \nonumber\\
 (\Delta m_D)_\Gamma &=& \frac{1}{\Gamma^2}|(V_{L}^{U})_{31}^*(V_{L}^{U})_{32}|^2 f_D^2 B_D \eta_D m_D, \nonumber\\
 (\Delta m_{B_d})_\Gamma &=& \frac{1}{\Gamma^2}|(V_{L}^{D})_{31}^*(V_{L}^{D})_{33}|^2 f_{B_d}^2 B_{B_d} \eta_{B_d} m_{B_d}, \nonumber\\ 
 \nonumber
 (\Delta m_{B_s})_\Gamma &=& \frac{1}{\Gamma^2}|(V_{L}^{D})_{32}^*(V_{L}^{D})_{33}|^2 f_{B_s}^2 B_{B_s} \eta_{B_s} m_{B_s}.\\
 \label{eq:FCNCmesons}
\end{eqnarray}
where $\Gamma= \sfrac{M_{\zp}}{\left|g^{\prime \prime}-g^{\prime} \right| }$ is a free parameter which has energy dimension, $B_K,B_D,B_{B_q}$ are the bag parameters, $f_K,f_D,f_{B_q}$ the decay constants, and $\eta_K,\eta_D,\eta_{B_q}$  the QCD leading order correction obtained in \cite{Misiak:1997ei,Barenboim:2000zz,Datta:2008qn,Dowdall:2019bea,Zyla:2020zbs,Branco:2021vhs,NguyenTuan:2020xls}, and $m_K,m_D,m_{B_d}$ the masses of the mesons. In {\it Table} \ref{tableI} we exhibit the values of these parameters.   

No sign of new physics has been observed in meson mixing. Thus, we use \eq{eq:FCNCmesons} to establish constraints on the effective energy scale $\Gamma$, by ensuring that the contribution from new physics is smaller than the experimental uncertainty reported in \tab{tableI}. For instance, having $(\Delta m_K)_{exp}=3.484\pm 0.006 \times 10^{-12}$~MeV, we require the new physics contribution to be smaller than $0.006 \times 10^{-12}$~MeV, and find the minimum value of $\Gamma$ that satisfies this criterion.
In summary, we consider the updated measurements of the neutral meson mixture bounds, and by evaluating the systematic errors, we ensure that the new physics is within the experimental error bars. In this way, our findings summarized in \tab{R_tab1} represent lower limits on the effective energy scale.

\begin{table}
    \centering
        \caption{Lower Mass bound on the effective energy scale $\Gamma$ rising from the meson mixing. See \eq{eq:FCNCmesons}.}
    \begin{tabular}{|c|c|c|c|c|}
    \hline 
    mixing   &  $K^0$    & $B_s$      & $B_d$     & $D^0$    \\
    \hline
    mixing 1 &  9.7 TeV  & 9.7 TeV    & 0.009 TeV & 815.4 TeV   \\
    \hline
    mixing 2 & 3.1 TeV   & 22.7 TeV   & 175.6 Te  & 1.3 TeV   \\
    \hline
    mixing 3 & 0.03 TeV  & 7.5 TeV    & 5.8 TeV   & 1.2 TeV  \\
    \hline
    mixing 4 & 0.3 TeV   & 118.1 TeV  & 3.4 TeV   & 0.1 TeV    \\
    \hline
    mixing 5 & 221.3 TeV   & 40.4 TeV   & 0.9 TeV   & 11.4  TeV     \\
    \hline
    \end{tabular}
    \label{R_tab1}
\end{table}

We have presented lower bounds on the effective energy scale $\Gamma$ for four meson systems, using five different mixings for the quark mixing matrices presented in the Appendix (refer to \cref{para1,para2,para3,para4,para5}). Our analysis indicates that the strongest constraint for the {\it mixing 1} arises from the $D^0-\overline{D^0}$ mixing, which requires $\Gamma > 815.4$ TeV. However, for {\it mixing 4}, the $D^0-\overline{D^0}$ system does not yield a strong constraint, while the $B_{s}^0-\overline{B_{s}^0}$ transition leads to $\Gamma > 118.1$~TeV.

Five different mixings were investigated to demonstrate the importance of considering all four meson mixings, rather than just one. We carefully selected values for the entries of the $V^{D}$ matrix and then derived the entries of the $V^{U}$ matrix to maintain consistency with the CKM matrix, as shown in \cref{VCKM}. Our analysis led us to the conclusion that $V^{D}_{33}$, $V^{D}_{31}$, and $V^{D}_{32}$ play a crucial role in the bounds derived from $B_q^0 \rightarrow{\overline{B_q^0}}$ ($q=d$ or $s$) and $K_0 \rightarrow{\overline{K}}_0$, respectively. On the other hand, the other entries in the $V^{D}$ matrix affect the $D_0 \rightarrow{\overline{D}}_0$ system. We emphasize that we kept the CKM matrix in agreement with the data throughout our analysis.


For {\it mixing 1}, with $D^0-\bar{D}^0$ mixing leading to $\Gamma > 815.4$~TeV. As one can see, it is important to consider all meson systems because they are sensitive to the mixing adopted in the quark mixing matrices. In \tab{R_tab1}, we report the limits found for the four meson mixings assuming five different mixings defined in the {\it \cref{A1}}. 

From \eq{eq:FCNCmesons}, it is evident that the individual mixing matrices play a crucial role in determining the FCNC amplitude. While maintaining the correct CKM matrix, one can manipulate these matrices to weaken or strengthen the bounds resulting from meson mixing. Thus, the study of the four meson systems is of paramount importance. It is possible to find a mixing where the $K^0-\overline{K^0}$ mixing leads to weak bounds. However, since any change in the entries of the mixing matrices for kaon mesons automatically affects the entries relevant for other mesons, it is essential to analyze all four meson systems to obtain reliable and accurate results.

We have concluded that finding a weak bound in a specific meson system does not necessarily mean the model is free from strong FCNC bounds. As shown in \cref{R_tab1}, the bounds derived from meson mixings significantly vary as we move from {\it mixing 1} to {\it 5}, with {\it mixings 1}, the strongest limit arises from the $D^0-\overline{D^0}$ mixture. In contrast, in {\it mixings 2, 4, and 5}, the strongest limits emerge from other meson systems.

The mixture labeled {\it mixings 3} provides weak constraints for all mesons. However, in other parameterizations, weaker constraints are observed for some mesons, but each provides a stronger limit for a specific meson mixture. As we have already mentioned in {\it mixings 1}, the strongest limit is observed for the $D^0-\overline{D^0}$ system, whereas in {\it mixings 2, 4, and 5}, the strongest limits are observed for $B_{d}^0-\overline{B_{d}^0}$, $B_{s}^0-\overline{B_{s}^0}$, and $K^0-\overline{K^0}$, respectively. However, {\it mixings 1, 4} provide the weakest constraints for $B_{d}^0-\overline{B_{d}^0}$ and $D^0-\overline{D^0}$. 
Therefore, it is crucial to consider all four meson systems when studying FCNC bounds to obtain reliable results. Thus far, we carried out our reasoning using an effective field theory. In the next section, we will adopt a concrete $\zp$ model.    

\section{\label{IV} FCNCs in \texorpdfstring{$\zp$}{$\zp$} model}

After discussing FCNCs using a simplified model, we use this knowledge to map our findings onto a concrete model that has a neutral current identical to Eq. \ref{eq:FCNC_new}. The model in question features an enlarged gauge sector, based on the $SU(3)_c \times SU(3)_L \times U(1)_N$ symmetry, it is known as the 3-3-1 model, for short. This model has a rich phenomenology in the context of dark matter \cite{Mizukoshi:2010ky,Alvares:2012qv,Profumo:2013sca}, muon anomalous magnetic moment \cite{Queiroz:2014zfa,deJesus:2020ngn}, neutrino masses \cite{Queiroz:2010rj}, collider physics \cite{CarcamoHernandez:2022fvl,Alves:2022hcp}, etc. The fermion generations are arranged as,

\begin{equation}
f^i_L=\left(
\begin{array}{c}
\nu^i_L\\
e^i_L\\
(\nu^c_R)^i
\end{array}\right) \sim (1,3,-1/3), \,\, e_R^i \sim (1,1,-1), 
\end{equation} 

\begin{equation}\begin{array}{l}
Q_{3L} = \left(\begin{array}{c}
u_3\\
d_3\\
u_{3}^\prime\end{array}\right)_L \sim (3, 3, 1/3),\\ \\
u_{3R} \sim (3, 1, 2/3),\
d_{3R} \sim (3, 1, -1/3),\ u^\prime_{1R} \sim (3, 1, 2/3),\\ \\
Q_{kL} = \left(\begin{array}{c}
d_k\\
- u_k\\
d_k^\prime\end{array}\right)_L \sim (3, \bar 3, 0), \\ \\
u_{kR} \sim (3, 1, 2/3),\ d_{kR} \sim (3, 1, -1/3),
d_{kR}^\prime \sim (3, 1, -1/3),\end{array} \end{equation}
where $i=1,\, 2,\, 3$,  indicate the three fermion generations, $k = 1, \, 2$, with $q^\prime = d_k^\prime, \, u_{3}^\prime$ being heavy exotic quarks with electric charges $Q(u^\prime_{3})=2/3$ and $Q(d^\prime_{1, \, 2})=-1/3$.

The effective lagrangians relevant for the meson mixings in the context of the 3-3-1 model are found to be \cite{GomezDumm:1994tz,Long:1999ij,Benavides:2009cn,CarcamoHernandez:2022fvl},

\begin{eqnarray}
 \mathcal{L}^{K_0-\bar{K}_0}_{\zp \, eff} &=& \Gamma^{\prime}\frac{M_Z^2}{M_{\zp}^2}|(V_{L}^{D})_{31}^*(V_{L}^{D})_{32}|^2|\bar{d}^{\prime}_{1L}\gamma_\mu d^{\prime}_{2L}|^2, \nonumber\\
 \mathcal{L}^{D_0-\bar{D}_0}_{\zp \, eff} &=& \Gamma^{\prime}\frac{M_Z^2}{M_{\zp}^2}|(V_{L}^{U})_{31}^*(V_{L}^{U})_{32}|^2|\bar{u}^{\prime}_{1L}\gamma_\mu u^{\prime}_{2L}|^2, \nonumber\\
 \mathcal{L}^{B^0_d-\bar{B}^0_d}_{\zp \, eff} &=& \Gamma^{\prime}\frac{M_Z^2}{M_{\zp}^2}|(V_{L}^{D})_{31}^*(V_{L}^{D})_{33}|^2|\bar{d}^{\prime}_{1L}\gamma_\mu d^{\prime}_{3L}|^2,\nonumber\\
 \mathcal{L}^{B^0_s-\bar{B}^0_s}_{\zp \, eff} &=& \Gamma^{\prime}\frac{M_Z^2}{M_{\zp}^2}|(V_{L}^{D})_{32}^*(V_{L}^{D})_{33}|^2 |\bar{d}^{\prime}_{2L}\gamma_\mu d^{\prime}_{3L}|^2,\nonumber
 \label{eq:FCNC8_2}
\end{eqnarray}and consequently,
\begin{eqnarray}
(\Delta m_K)_{\zp} &=& \Gamma^{\prime} \frac{M_Z^2}{M_{\zp}^2} |(V_{L}^{D})_{31}^*(V_{L}^{D})_{32}|^2 f_K^2 B_K \eta_K m_K, \nonumber\\
 (\Delta m_D)_{\zp} &=& \Gamma^{\prime} \frac{M_Z^2}{M_{\zp}^2}|(V_{L}^{U})_{31}^*(V_{L}^{U})_{32}|^2 f_D^2 B_D \eta_D m_D, \nonumber\\
 (\Delta m_{B_d})_{\zp} &=& \Gamma^{\prime} \frac{M_Z^2}{M_{\zp}^2}|(V_{L}^{D})_{31}^*(V_{L}^{D})_{33}|^2 f_{B_d}^2 B_{B_d} \eta_{B_d} m_{B_d}, \nonumber\\ 
 \nonumber
 (\Delta m_{B_s})_{\zp} &=& \Gamma^{\prime} \frac{M_Z^2}{M_{\zp}^2}|(V_{L}^{D})_{32}^*(V_{L}^{D})_{33}|^2 f_{B_s}^2 B_{B_s} \eta_{B_s} m_{B_s}.\\
 \label{eq:FCNC3_2}
\end{eqnarray}

where $\Gamma^{\prime}= \frac{4 \sqrt{2}G_F C_W^4}{3-4 S_{W}^2}$, with $G_F$ being the Fermi constant. One can easily notice that $\Gamma^\prime$ has energy dimension $E^{-2}$.

\begin{eqnarray}
\label{zprimau-1_2}
\ensuremath{\mathcal{L}}^{\zp}_{u} & = & \frac{g}{2C_{W}}\left(\frac{(3-4S_{W}^{2})}{3\sqrt{3-4S_{W}^{2}}}\right) \left[\bar{u}_{kL}\gamma_{\mu}u_{kL}\right] \zp_{\mu} \nonumber \\
                                          &   & - \frac{g}{2C_{W}}\left(\frac{(3-2S_{W}^{2})}{3\sqrt{3-4S_{W}^{2}}}\right)\left[\bar{u}_{3L}\gamma_{\mu}u_{3L}\right] \zp_{\mu},\\
\ensuremath{\mathcal{L}}^{\zp}_{d} & = & \frac{g}{2C_{W}}\left(\frac{(3-4S_{W}^{2})}{3\sqrt{3-4S_{W}^{2}}}\right)
\left[\bar{d}_{kL}\gamma_{\mu}d_{kL}\right] Z_{\mu}^{\prime}  \nonumber \\
                                          &   & - \frac{g}{2C_{W}}\left(\frac{(3-2S_{W}^{2})}{3\sqrt{3-4S_{W}^{2}}}\right)
\left[\bar{d}_{3L}\gamma_{\mu}d_{3L}\right] Z_{\mu}^{\prime},
\label{eq:FCNC1_2} 
\end{eqnarray} 
with $k=1,\, 2$, indicate the generation indices, and $C_{W} \equiv cos\theta_{W}$, $S_{W} \equiv sin\theta_{W}$, with $\theta_{W}$ being the Weinberg angle.

\section{\label{V}Discussions}

\begin{figure}
 \centering
 \includegraphics[width=0.98\columnwidth]{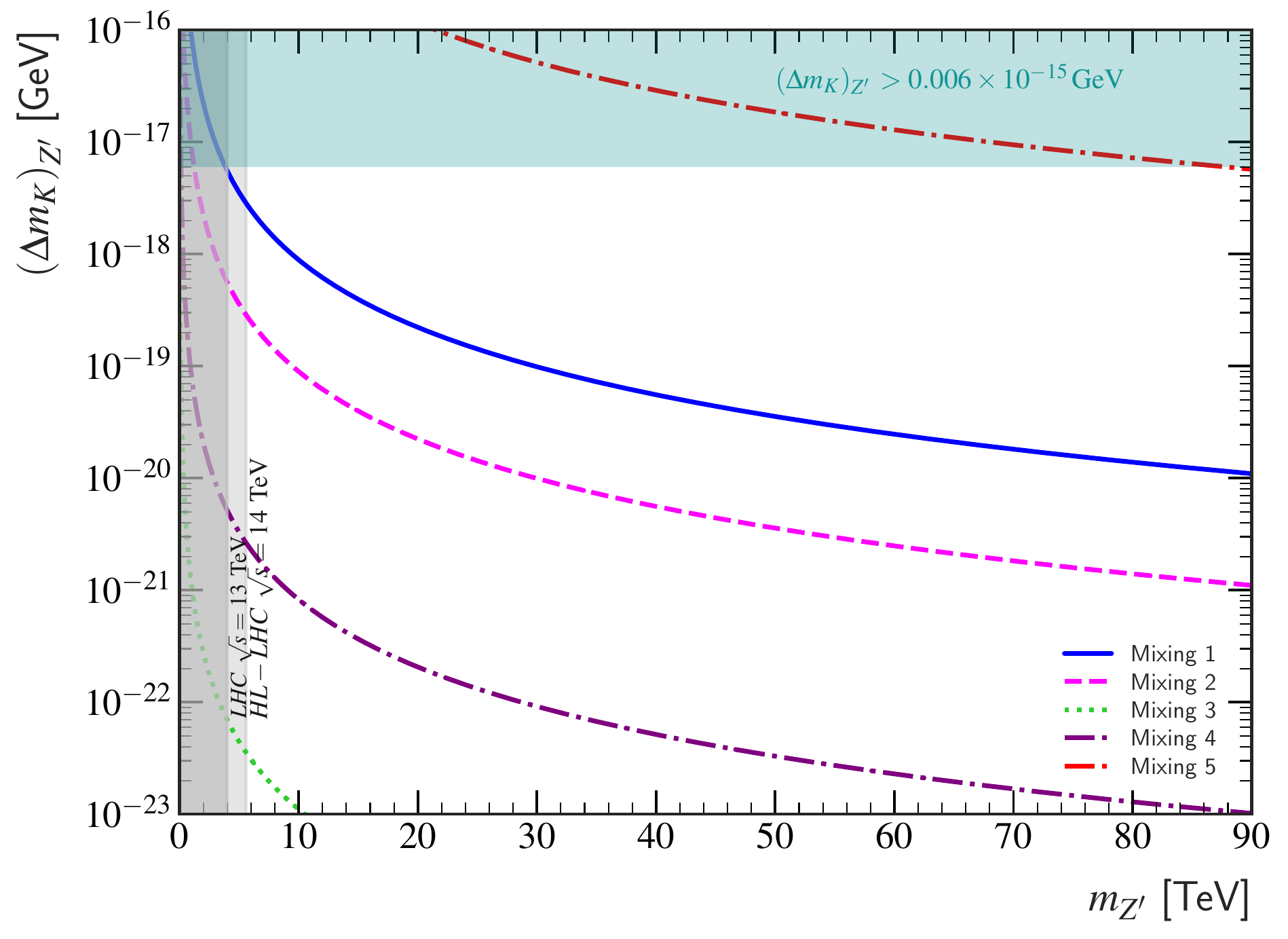}
 \caption{The solid blue, dashed magenta, dash-dot violet and red, and dotted green lines correspond to $\zp$ contribution to the $(\Delta m_K)_{\zp}$ as a function of it mass (\eq{eq:FCNC3_2}), for the five mixings of the $V_{L}^{D}$ matrix, see \cref{para1,para2,para3,para4,para5}. The light cyan region corresponds to the FCNC exclusion region. We overlaid current and projected colliders bounds (gray regions). }
 \label{finalplot1}
\end{figure}

\begin{figure}
 \centering
 \includegraphics[width=0.98\columnwidth]{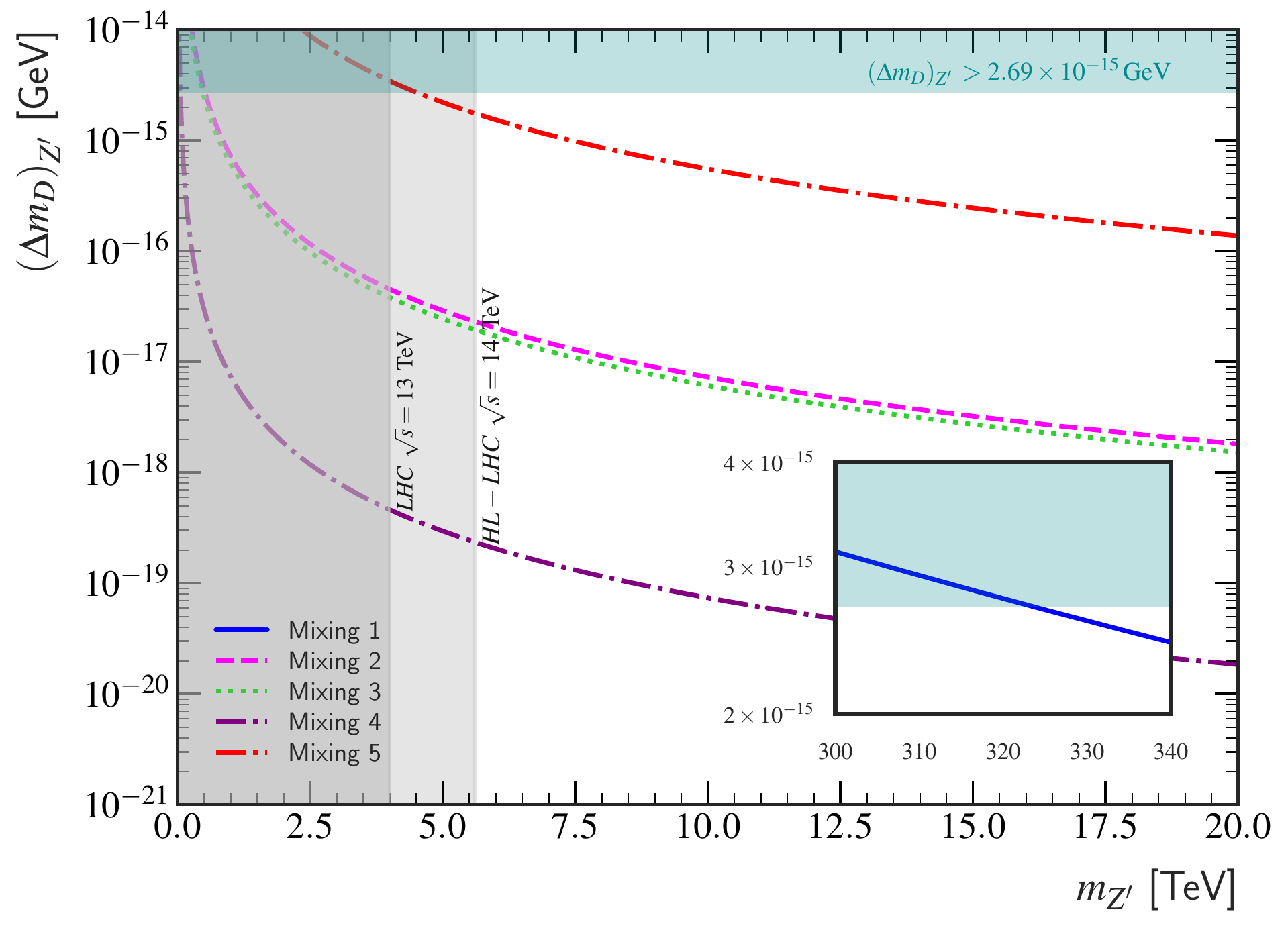}
 \caption{The solid blue, dashed magenta, dash-dot violet and red, and dotted green lines correspond to $\zp$ contribution to the $(\Delta m_D)_{\zp}$ as a function of it mass (see \eq{eq:FCNC3_2}), for the five mixings of the $V_{L}^{D}$ matrix discussed in this work. The light cyan region corresponds to the FCNC exclusion region. The gray regions represent the current and projected colliders bounds. The subfigure located within the main figure in the lower right corner corresponds to the Mixing 5, which is not visible in the main plot due to the plotted range.}
  \label{finalplot2}
\end{figure}

\begin{figure}
 \centering
 \includegraphics[width=0.98\columnwidth]{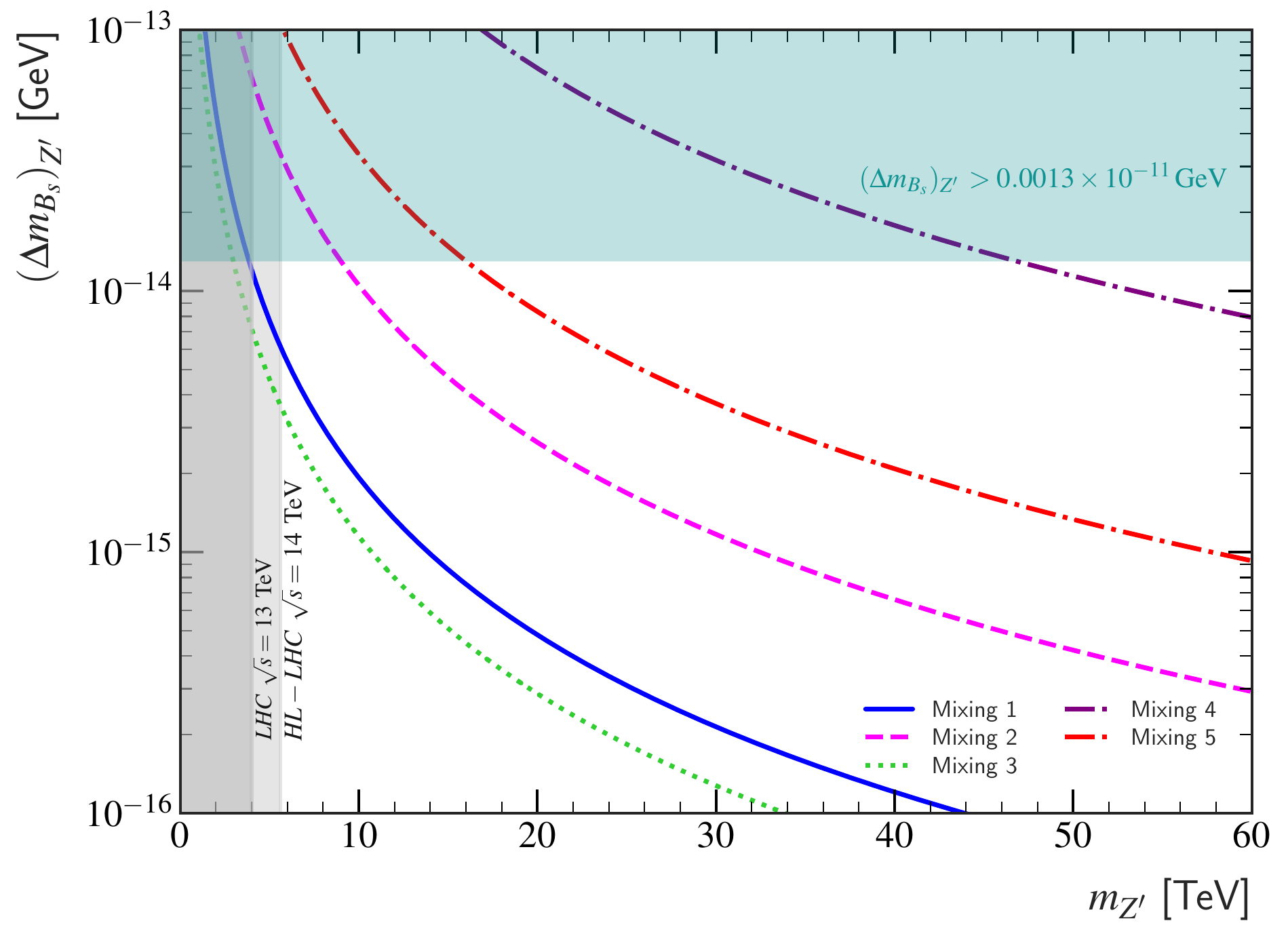}
 \caption{Similarly to Fig.\ref{finalplot2} but for the $B_s^-\bar{B_s}^0$ system.}
  \label{finalplot3}
\end{figure}

\begin{figure}
 \centering
 \includegraphics[width=0.98\columnwidth]{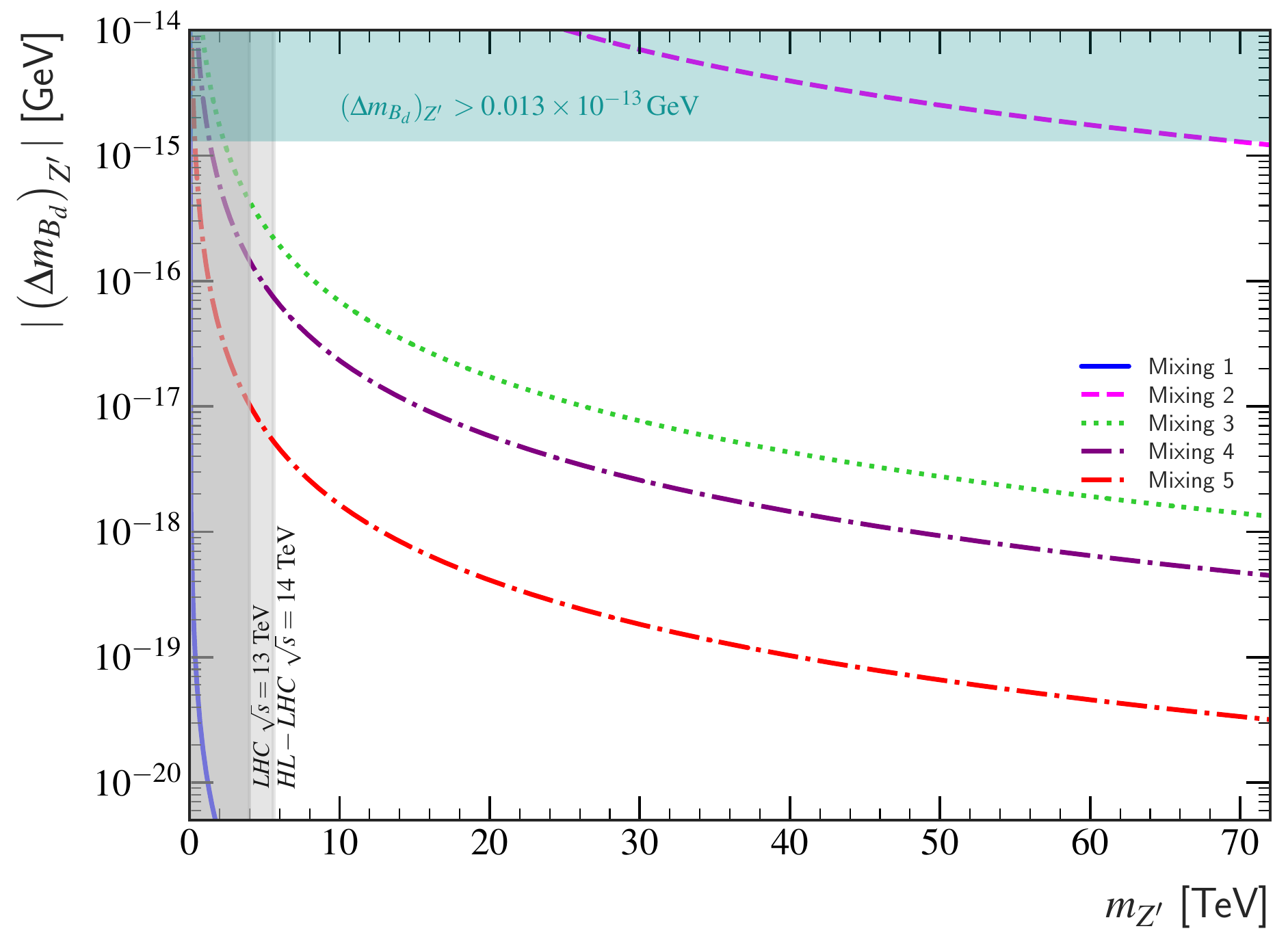}
 \caption{Similarly to Fig.\ref{finalplot3} but for the  $B_d^0-\bar{B_d}^0$ mixing}.
  \label{finalplot4}
\end{figure}

As before, we set values for entries of the $V^{D}$ matrix and derive those from the $V^{U}$ matrix to preserve the CKM matrix, \cref{VCKM}. Analogously, adopting five different mixings \cref{para1,para2,para3,para4,para5}, we derive lower bounds, but this time on the mass of the $\zp$ field. Our findings are summarized in \tab{R_tab2}. We plotted them in \cref{finalplot1,finalplot2,finalplot3,finalplot4} respectively. 
In addition to the lower mass bounds derived from FCNC, we overlay collider bounds on the $\zp$ which were derived using LHC data, and projections for HL-LHC according to \cite{Alves:2022hcp}. In particular, it was found that LHC data yields $M_{\zp} \geq 4$~TeV, while HL-LHC, in case of no signal, is expected to impose
$M_{\zp} \geq 5.6 $~TeV. Those are the bounds superimposed in the figures.

\begin{table}
    \centering
        \caption{Lower Mass bound on the $m_{\zp}$ obtained in the four meson systems for the 3-3-1 Model. See \eq{eq:FCNCmesons}.}
    \begin{tabular}{|c|c|c|c|c|}
    \hline 
    mixing   &  $K_0$    & $B_s^0$   & $B_d^0$    & $D_0$     \\
    \hline
    mixing 1 &  3.8 TeV  & 3.9 TeV   & 0.003 TeV    & 323.5 TeV    \\
    \hline
    mixing 2 &  1.2 TeV  & 9.0 TeV   & 69.7 TeV    & 0.5 TeV    \\
    \hline
    mixing 3 &  0.01 TeV & 3.0 TeV   & 2.3 TeV    & 0.5 TeV   \\
    \hline
    mixing 4 &  0.1 TeV   & 46.9 TeV   & 1.3 TeV   & 0.05 TeV    \\
    \hline
    mixing 5 &  87.8 TeV  & 16.0 TeV   & 0.4 TeV     & 4.5 TeV      \\
    \hline
    \end{tabular}
    \label{R_tab2}
\end{table}

In the graphs \cref{finalplot1,finalplot2,finalplot3,finalplot4}, you can observe different colored lines representing $\zp$ contribution to the $(\Delta m_j)_{\zp}$ in four meson systems ($j = K, D, B_S, B_d$) as mentioned in \eq{eq:FCNC3_2}. The solid blue, dashed magenta, dash-dot violet, and red, and dotted limegreen lines correspond to the different meson systems. The light cyan regions mark the excluded regions by FCNC measurements while the dark gray parameter space is excluded by LHC data. The projected HL-LHC exclusion bound is also displayed within a light gray area.

Putting our results into perspective with LHC data and the projected HL-LHC bound, we conclude that if we adopted {\it mixing 3} for the quark mixing matrices, the lower mass bounds stemming FCNCs are subdominant compared to colliders. We stress that this conclusion goes opposite to the common statement, which states that FCNCs probe energy scales beyond high-energy colliders.
However, for {\it mixings 1, 2, 4, and 5}, we choose four different scenarios in which each meson system yields stringent bounds. We emphasize that in our analysis, we kept the CKM matrix in agreement with the data. 

In {\it mixing 1}, the strongest constraints come from the $D_0$ mixing, whereas for {\it mixing 2} the $B_{d}$ system is more constraining, for {\it mixing 4} is the $B_{d}$, and lastly, for {\it mixing 5} the $K$. 

We have shown that this statement that FCNCs probe much higher energy scales than colliders unless a highly non-trivial flavor structure is present needs to be revisited. We have explicitly proved this in a model in which this natative has also been assumed. In summary, FCNC does not necessarily lead to constraints stronger than colliders, strengthening the need and importance for future colliders such as HL-LHC.


\section{\label{VI}Conclusion}

We have conducted a thorough investigation on the influence of new physics processes on the meson mixings ($B_d^0 \rightarrow{\overline{B_d^0}}$, $B_s^0 \rightarrow{\overline{B_s^0}}$, $K_0 \rightarrow{\overline{K}}_0$ and $D_0 \rightarrow{\overline{D}}_0$) utilizing effective operators and a $\zp$ model embedded in a $SU(3)_C\otimes SU(2)_L \otimes U(1)_N$ gauge group. Since the meson mixings are sensitive to different entries of the quark mixing matrices, we assessed the impact of the entries adopted in the quark mixing matrices on the results. Our findings reveal that these mixings can significantly affect the size of the new physics contribution to meson mixing and the lower bounds on the new physics energy scale or the mass of the $\zp$ boson.

Furthermore, we observed that the lower bounds on the $\zp$ mass or the new physics energy scale can be substantially weakened for certain mixings. Specifically, we have conservatively determined the lower bounds on the $\zp$ mass to be $m_{\zp}>3.0$~TeV for {\it mixing 3}, which are much weaker than those previously reported in the literature. Our results demonstrate that the LHC and HL-LHC can potentially probe $\zp$ masses larger than those excluded through FCNC studies. Therefore, the LHC and HL-LHC will constrain flavor-changing neutral processes.

\acknowledgments
This work was financially supported by Simons Foundation (Award Number:1023171-RC), FAPESP Grant
2021/01089-1, ICTP-SAIFR FAPESP Grants 2021/14335-0, CNPq Grant 307130/2021-5, FONDECYT Grant 1191103 (Chile)  and ANID-Programa Milenio-code ICN2019\_044. 
FSQ thanks Universidad Técnica Frederico Santa Maria for the hospitality during the initial stage of this work. Y.S.V. expresses gratitude to São Paulo Research Foundation (FAPESP) under Grant No. 2018/25225-9 and 2023/01197-4. Y.M.O.T. acknowledges financial support from CAPES under grants 88887.485509 / 2020-00. S.K. is supported by ANID PIA/APOYO AFB180002 (Chile) and 
by ANID FONDECYT (Chile) No. 1230160. TBM acknowledges ANID - Millennium Program - ICN2019 044 and ANID-Chile FONDECYT (grant No. 3220454) for financial support.
A.Z. is supported by Fondecyt Grant 1230110, ANID - Millennium Program - ICN2019 044 and o ANID PIA/APOYO AFB180002 (Chile).

\appendix

\section{Choices of the mixing matrices \label{A1}}
We will assume two different choices of the mixing matrices yield significant changes in the new physics contribution to the mass difference systems. In this way, we can assess the impact of such mixings. 

We adopt  {\it mixing 1},

\begin{align}
\label{para1}
V_{L}^{D}& =  \nonumber \\
&\left( \begin{array}{ccc}
-0.3519353880 & 0.0024752316 & 0.9360210230 \\
-0.9360242340 & -0.0005671345 & -0.3519350960 \\
-0.0003402711 & -0.9999967760 & 0.0025164715
\end{array} \right) \nonumber  \\
V_{L}^{U}&=  \nonumber\\
&\left( \begin{array}{ccc}
-0.3388974011 & -0.0375821790 & 0.9322810984 \\
-0.9134414582 & -0.2257444347 & -0.3596697258 \\
-0.2253215319 & -0.9734581356 & -0.0385885317
\end{array} \right), 
\end{align}

 {\it mixing 2},
\begin{align}
\label{para2}
V_{L}^{D}& =  \nonumber \\
&\left( \begin{array}{ccc}
-0.9800000000 & 0.0717338200 & -0.0224212200 \\
-0.0711610200 & -0.9971419100 & -0.0253794100 \\
-0.0180000000 & -0.0060000000 & 0.9800000000
\end{array} \right) \nonumber  \\
V_{L}^{U}&=  \nonumber\\
&\left( \begin{array}{ccc}
-0.9723022675 & -0.1579745355 & -0.0238748546 \\
-0.2954345497 & -0.9877260050 & -0.0490589982 \\
-0.0261307731 & 0.0134797091 & 1.0176914369
\end{array} \right), 
\end{align}

 {\it mixing 3},
\begin{align}
\label{para3}
V_{L}^{D}& =  \nonumber \\
&\left( \begin{array}{ccc}
-0.0334508531 & 0.9993711310 & -0.0117636070 \\
-0.9994401830 & -0.0354556072 & -0.0002075334 \\
-0.0006009615 & 0.0020017501 & 0.9699307850
\end{array} \right) \nonumber  \\
V_{L}^{U}&=  \nonumber\\
&\left( \begin{array}{ccc}
0.1903163758 & 0.9648639798 & 0.0386895011 \\
-0.9812691235 & -0.2573124820 & -0.0104893953 \\
0.0059026489 & 0.0547635775 & 1.0305590529
\end{array} \right), 
\end{align}

 {\it mixing 4},
\begin{align}
\label{para4}
V_{L}^{D}& =  \nonumber \\
&\left( \begin{array}{ccc}
-0.0768067292 & 0.9964491790 & 0.0344928899 \\
-0.9970459430 & -0.0767490000 & -0.0029965491 \\
-0.0003386141 & -0.0306211511 & 0.9994004510
\end{array} \right) \nonumber  \\
V_{L}^{U}&=  \nonumber\\
&\left( \begin{array}{ccc}
0.1495093747 & 0.9541730910 & 0.0707835338 \\
-0.9887526569 & -0.2990329892 & -0.0143868278 \\
-0.0044325138 & 0.0080165519 & 0.9972312495
\end{array} \right), 
\end{align}

{and  \it mixing 5},
\begin{align}
\label{para5}
V_{L}^{D}& =  \nonumber \\
&\left( \begin{array}{ccc}
-0.7485855660 & 0.0209062160 & 0.6627085190 \\
-0.6630381690 & -0.3731665020 & -0.7422237300 \\
-0.0081782900 & -0.9500000000 & 0.0110161269
\end{array} \right) \nonumber  \\
V_{L}^{U}&=  \nonumber\\
&\left( \begin{array}{ccc}
-0.7243902225 & -0.1270248872 & 0.6557696276 \\
-0.6517094009 & -0.1839389342 & -0.7566148098 \\
0.0032124832 & -0.9552692222 & 0.2683715404
\end{array} \right), 
\end{align}

Knowing the entries of the up-quark and down-quark mixing matrices $V^u_L$ and $V^d_L$, we determine the $\zp$ contribution to the mass difference of the meson systems and consequently place a lower mass bound. We adopt these mixings because they yield the strongest and weakest 3-3-1 contributions to FCNC processes while keeping the CKM matrix in agreement with the data. It is important to mention that the Cabibbo-Kobayashi-Maskawa (CKM) matrix is not perfectly unitary based on experimental values. This means that when one of the matrices is fixed to be unitary (for example, $V^{D}_{L}$), the other does not turn out to be completely unitary. However, its characteristics remain similar to those of the CKM matrix. With this information at hand, we use \eqs{eq:FCNCmesons} and \eqref{eq:FCNC3_2} to plot our findings in \figs{finalplot1,finalplot2,finalplot3,finalplot4}.

\def\bibsection{\section*{References}}
\bibliographystyle{utphys}
\bibliography{darkmatter}

\end{document}